\documentclass[fleqn]{article}
\usepackage{emlines2,bezier,amscd}
\usepackage{amssymb}
\newcommand{\df}{\ensuremath{\mathrm{DEF}_{\mathcal{M}}(\mathcal{B})}}
\newcommand{\de}[2]{\ensuremath{(\iota_{#1},\iota_{#2})}}
\newcommand{\depr}[2]{\ensuremath{(\iota'_{#1},\iota'_{#2})}}
\newcommand{\tm}[2]{\ensuremath{#1\times #2}}

\newcommand{\Diff}[1]{\ensuremath{{\rm Diff}(#1)}}
\newcommand{\pr}[1]{\ensuremath{{\rm pr}_{#1}}}
\newcommand{\hm}[1]{\ensuremath{\mathrm{H}_{#1}}}
\newcommand{\hmt}[2]{\ensuremath{#1\stackrel{\mathrm{H}}{\sim} #2}}

\newcommand{\mt}[1]{\ensuremath{\mathrm{MOT}_{\mathcal{M}}(#1)}}
\newtheorem{pred}{Proposition}

\title{d-OBJECTS KINEMATICS ON SMOOTH MANIFOLDS}
\author{Sergey S.Kokarev\thanks{sergey@yspu.yar.ru}}
\date{Department of theoretical physics, r.409, YSPU,
                       Respublikanskaya 108, Yaroslavl,150000, Russia}

\begin{document}
\twocolumn[\maketitle

\begin{abstract}
The kinematical  part of general theory of deformational structures on smooth
manifolds is developed. We introduce general concept of d-objects deformation,
then within the set of all such deformations we develop some special algebra
and investigate group and homotopical properties of the set.  In case of proper
deformations  some propositions, generalizing isometry theory on Riemannian
manifolds are formulated.
    \end{abstract}
\vspace{1cm}]

\section{Introduction}

Recent time the strong tendency towards the inclusion of  an embedded objects
into the scope of  theoretical and mathematical physics is observed (see
references in \cite{pav}). One should refer to the subject all string and
$p-$brane models \cite{green,plectures}, including their supersymmetric and
noncommutative generalizations \cite{noncomm}, embedding methods of GR and its
alternative formulations and generalizations \cite{kobayasi}, geometrical
methods of nonlinear differential equations theory and jets approach
\cite{vinogr} and many other things. Probably, this central position of the
"embedded objects" in modern physics can't be accidental:  it may reflect
either multidimensional nature of physical reality, observed through all its
levels, or some "immanent" to us, as observers, ways of its descriptions.

On lagrangian level, within variety of our field theoretical models, exploiting
embedded objects, the\-re easily can be seen its amazing and, in our opinion,
deep interrelation with  elasticity theory of continious media \cite{land}, may
be with general properties such as nonlinearity, plasticity, viscosity,
anisotropy, and may be possessing internal spin, nematic or smectic structures
or memory \cite{saharov,tar,hehl,dmitriev}. Particularly, in the papers
\cite{kok1,kok2,kok3,kok4} it has been shown, that Einstein GR and standard
classical solid dynamics admit natural formulation in terms of mechanical
straining of thin 4D plates and 4D strings (strongly tensed bars)
correspondingly.

The interesting and very important problem, arising now, is to extract and
formulate general ideas of continuous media physics in its the most abstract
and general form, independent on peculiarities of one or another theory,  both
to be able to see and use it within general context of our field theoretical
models and to apply it successfully in some concrete situations. So, we intend
to follow the line of investigations, that will be called here {\it general
theory of deformational structures,} the aim of which --- to formulate and work
out universal language for the objects, that are able, in some sense, to be
"deformed".

In the present short paper we give the  sketch of the approach, and directly
generalize standard elasticity theory for bodies in $\mathbb{E}^3$ for the case
of arbitrary smooth manifold and smooth forms as deformational metrics (below
$d-$manifold, $d-$objects and  $d-$metrics). We take here for consideration
only {\it kinematics}, which is totally described by the {\it free
deformational structure}. Dynamical deformational structure consist of free one
and variational procedure, which we'll not consider in the paper. By the volume
restrictions, all statements are only formulated (without proofs). Anywhere, if
it is possible, we use standard notations of smooth manifold theory, as, for
example, in \cite{warner}.

\section{Definitions}

We call {\it free deformational structure  $\mathfrak{D}_0$} the collection
$\langle\mathcal{B},\mathcal{M},\mathcal{E},\Theta\rangle,$ where:

$\mathcal{B}$ and $\mathcal{M}$ --- smooth, connected, closed manifolds, ${\rm
dim}\,\mathcal{B}=d,$ ${\rm dim}\,\mathcal{M}=n\ge d;$

$\mathcal{E}$ --- set of all possible smooth embeddings
$\mathcal{B}\hookrightarrow\mathcal{M};$

$\Theta\in\Omega^{\otimes p}(\mathcal{M})$ --- some smooth real form of rank
$p$ on $\mathcal{M}.$ In what follows we shall use the abbreviations:
$\mathcal{B}$
--- {\it $d-$body},  $\mathcal{M}$ --- {\it $d$-manifold,}
$\Theta$
--- {\it $d$-metrics,} and image
\begin{equation}\label{emb}
\iota(\mathcal{B})\equiv\mathcal{S}\subset\mathcal{M}
\end{equation}
for some $\iota\in\mathcal{E}$ --- {\it $d$-objects} or {\it deformant.}

Embedding $\iota$ induces form $(d\iota)^\ast\Theta$ on
$T^\ast\mathcal{B}^{\otimes p},$ where $(d\iota)^\ast$ --- embedding $\iota$
codifferential,  acting in tensor product $T^\ast\mathcal{S}^{\otimes p}.$ Let
consider some another embedding
$\iota'\in\mathcal{E},$
which induces its own  $d-$object
$\iota'(\mathcal{B})\equiv\mathcal{S'}\subset\mathcal{M}.$ In
$T^\ast\mathcal{B}^{\otimes p}$ we'll have form $(d\iota')^\ast\Theta.$ Easily
to see, that composition
\begin{equation}\label{def}
\iota'\circ\iota^{-1}\equiv\zeta
\end{equation} is diffeomorphism
$\mathcal{S}\to\mathcal{S'}=\zeta(\mathcal{S}),$  which we'll call {\it
deformation} of $d-$object in $\mathcal{M}$.

The deformation $\zeta$ has natural local measure
--- difference of two forms, taken in the same point of $b\in\mathcal{B}$:
\begin{equation}\label{formb}
(d\iota')^\ast\Theta_{\iota'(b)}-(d\iota)^\ast\Theta_{\iota(b)}\equiv
\Delta_{\mathcal{B}}(b),
\end{equation}
where we have introduced notation $\Delta_{\mathcal{B}}$ for {\it deformation
form} on $\mathcal{B}.$ Using definition (\ref{def}) of deformation and well
known property of codifferential: $d(\alpha\circ\beta)^\ast=(d\beta)^\ast\circ
(d\alpha)^\ast$ one can  obtain the equivalent representation:
\begin{equation}\label{formb1}
\Delta_{\mathcal{B}}=(d\iota)^\ast ((d\zeta)^\ast \Theta-\Theta),
\end{equation}
and define deformation form
\begin{equation}\label{forms}
\Delta_{\mathcal{S}}\equiv
((d\iota)^\ast)^{-1}\Delta_{\mathcal{B}}=(d\zeta)^\ast\Theta-\Theta
\end{equation}
on deformant $\mathcal{S}.$ Lets note, that representations (\ref{formb1}) and
(\ref{forms}) correspond to material and referent descriptions of deformable
bodies configurations in classical dynamics \cite{trus}. All is illustrated on
diagrams (\ref{diag1}).
\begin{equation}\label{diag1}
\begin{CD}
\mathcal{B} @>\iota'>>\mathcal{S'} @. \subseteq\mathcal{M}\\ @|
@A\zeta=\iota'\circ\,\iota^{-1}AA@.\\ \mathcal{B} @>\iota>>\mathcal{S}@.
\subseteq\mathcal{M}
\end{CD}
\hspace{0.5cm}
\begin{CD}
T^\ast\mathcal{B}^{\otimes p} @<(d\iota')^\ast<<T^\ast\mathcal{S'}^{\otimes p
}@.{}\\ @| @V(d\zeta)^\ast VV@.\\ T^\ast\mathcal{B}^{\otimes p}
@<(d\iota)^\ast<<T^\ast\mathcal{S}^{\otimes p}@. {}@.
\end{CD}
\end{equation}

\section{Pseudogroup of deformations of $d-$body in $\mathcal{M}$.}

As it follows from definition (\ref{def}), the set of all defromations of
$d-$objects in $\mathcal{M}$, which we'll denote
$\mathrm{DEF}_{\mathcal{M}}(\mathcal{B}),$ can be treated as image of the
surjective map $\phi:\ \mathcal{E}\times\mathcal{E}\to\df,$ acting by the rule:
\begin{equation}\label{rule}
\phi(\iota_\alpha,\iota_\beta)=\iota_\beta\circ\iota_\alpha^{-1}\equiv\zeta_{\alpha\beta}.
\end{equation}
The following proposition clears the relation between
$\tm{\mathcal{E}}{\mathcal{E}}$ and $\df.$
\begin{pred}\label{corresp}\ Fibre
$\phi^{-1}(\zeta)=\{d\in\mathcal{E}\times\mathcal{E}\,|\ d=(\iota_\zeta\circ
l,\zeta\circ\iota_\zeta\circ l)\},$ where $l$ runs all elements from the set
${\rm Diff}(\mathcal{B})$ --- group of diffeomorphisms of  $\mathcal{B},$
$\zeta$
--- some element of $\df,$ and embedding $\iota_\zeta$ satisfies the condition:
${\rm Im}(\iota_\zeta)={\rm Dom}(\zeta).$
\end{pred}

Note, that the set $\Diff{\mathcal{B}}$ can be viewed as subgroup of ${\rm
Aut}(\mathcal{E}),$ since $\mathcal{E}\circ\Diff{\mathcal{B}}=\mathcal{E}.$ Now
we claim, that $\df$ is factor-set
$\pi_D(\tm{\mathcal{E}}{\mathcal{E}})\equiv\widehat{\tm{\mathcal{E}}{\mathcal{E}}},$
consisting of classes  of elements $[\de{\alpha}{\beta}]_D\equiv
d_{\alpha\beta},$ that lie in the same fibre, where  the element
$\de{\alpha}{\beta}\in\tm{\mathcal{E}}{\mathcal{E}}$ does. Here $\pi_D$
--- factorization map by equivalence $\stackrel{D}{\sim},$
given by division $\tm{\mathcal{E}}{\mathcal{E}}$ by fibres $\phi^{-1}(\zeta).$

Since $\widehat{\tm{\mathcal{E}}{\mathcal{E}}}$ is not direct product, then
projections ${\rm pr}_1,\ {\rm pr}_2$ don't defined on
$\widehat{\tm{\mathcal{E}}{\mathcal{E}}}$ in common sense. Let ${\rm
Emb}_{\mathcal{M}}(\mathcal{B})$ be the set of all images $\{{\rm
Im}(\iota)\}_{\iota\in\mathcal{E}}.$ In what follows we'll use projections as
mappings ${\rm pr}_i:\ \widehat{\tm{\mathcal{E}}{\mathcal{E}}}\to{\rm
Emb}_{\mathcal{M}}(\mathcal{B}) $ $i=1,2,$ which acts on every class
$[(\iota_1,\iota_2)]_D$ by the rule:
\begin{equation}\label{proj}
{\rm pr}_1[(\iota_1,\iota_2)]_D\equiv\iota_1(\mathcal{B});\ \ {\rm
pr}_2[(\iota_1,\iota_2)]_D\equiv\iota_2(\mathcal{B}).
\end{equation}
The formal notation ${\rm pr}_i(d)$ will be understood below exactly in this
sense.

 On the set $\widehat{\tm{\mathcal{E}}{\mathcal{E}}}$ one can introduce
binary relation:
\begin{equation}\label{turnir}
\rho=\{(d_1,d_2)\in\tm{\widehat{\tm{\mathcal{E}}{\mathcal{E}}}}{\widehat{\tm{\mathcal{E}}{\mathcal{E}}}}\,|\
\pr{1}(d_1)=\pr{2}(d_2)\},
\end{equation}
It is easily to check, that $\rho$ is $\rm T-$reflective  and $\rm
T-$antisymmetric, i.e.: $$(d,d^{\rm T})\in\rho,\ \ $$ and, if simultaneously
$$(d_{1},d_{2})\in\rho\ \mbox{\rm и}\ (d_{2},d_{1})\in\rho,\ \mbox{\rm then}\
d_2=d_1^{\rm T}.$$ Here $(d^{\rm T})_{\alpha\beta}\equiv d_{\beta\alpha}.$
We'll call this relation {\it $\rm T-$tournament\footnote{Tournament is
reflective and antisymmetric binary relation.}.}

Let $d\in \widehat{\tm{\mathcal{E}}{\mathcal{E}}}.$ Let denote as $Y^{\mp}_d$
the following subsets:
$$Y^{-}_d\equiv\{d'\in\widehat{\tm{\mathcal{E}}{\mathcal{E}}}\,|\
(d,d')\in\rho\};$$ $$
Y^{+}_d\equiv\{d'\in\widehat{\tm{\mathcal{E}}{\mathcal{E}}}\,|\
(d',d)\in\rho\}, $$ which will be referred to as  the {\it bottom } and {\it
top} sets of the element $d$ respectively.

\begin{pred}\label{ps}
On the set  $\widehat{\tm{\mathcal{E}}{\mathcal{E}}}$ with  $\rm
T-$tour\-na\-ment $\rho$ one can define pseudogroup structures \footnote{Let
remind, that pseudogroup is a set of elements $\mathcal{A}$, where group
multiplication $\ast$ is defined may be on some subset (binary relation)
$\mathcal{U}\subset\tm{\mathcal{A}}{\mathcal{A}}$ and where hold the following
properties: associativity,  for every $a\in\mathcal{A}$ there exist unique
right $e_a^{-}$ and left $e_a^{+}$ units elements (generally speaking depending
on $a$), lied in $\mathcal{A}$ and there exists unique inverse element
$a^{-1},$ lying in $\mathcal{A},$ such that $a\ast e^{-}_a=e^{+}_a\ast a=a$ and
$a\ast a^{-1}=e^{+}_a,\ \ a^{-1}\ast a=e^{-}_a.$}.
\end{pred}

It is easily to check in our case, that pseudogroup  operation can be defined
as follows. For any $d\in\widehat{\tm{\mathcal{E}}{\mathcal{E}}}$ and for all
$d'\in Y^{-}_d$ и $d''\in Y^{+}_d$ we define multiplication  "$\ast$" from the
right and from the left by the rules: $$d\ast d'\equiv d\circ d';\ \ d''\ast
d\equiv d''\circ d.$$ In components: $$d\ast
d'=[\de{1}{2}]_D\ast[(\iota',\iota_1)]_D\equiv[(\iota',\iota_2)]_D;$$ $$d''\ast
d= [(\iota_2,\iota'')]_D\ast[\de{1}{2}]_D\equiv[(\iota_1,\iota'')]_D.$$ Unit
elements will be given by the expressions:
$$e^{-}_d\equiv[(\pr{1}(d),\pr{1}(d))]_D\in
\Delta(\widehat{\tm{\mathcal{E}}{\mathcal{E}}}),$$
$$e^{+}_d\equiv[(\pr{2}(d),\pr{2}(d))]_D\in
\Delta(\widehat{\tm{\mathcal{E}}{\mathcal{E}}}),$$ where
$\Delta(\widehat{\tm{\mathcal{E}}{\mathcal{E}}})$ --- diagonal of
$\widehat{\tm{\mathcal{E}}{\mathcal{E}}}$ (in the sense (\ref{proj})) Also, for
every $d\in\widehat{\tm{\mathcal{E}}{\mathcal{E}}}$ there exist unique unit
element  $d^{-1}$ and it easily to check in components, that $d^{-1}=d^{\rm
T}.$

So, {\it the set of deformations $\df$ --- pseudogroup.}

\section{Homotopies and special deformations}

Lets consider $\hm{\mathcal{E}},$ consisting of homotopic classes of embeddings
$\mathcal{E}.$ We define here {\it strong smooth homotopy} of embedding
$\iota\in\mathcal{E}$ as smooth mapping $F:\
\tm{\mathcal{B}}{I}\to\mathcal{M},$ where $I=[0,1],$ such, that
$F(\mathcal{B},0)=\iota$ and $F(\mathcal{B},s)\equiv
F_s(\mathcal{B})\in\mathcal{E}$ for every $s\in I.$ Two embeddings $\iota$ and
$\iota'$ are said to be homotopic: $\hmt{\iota}{\iota'}$, if there exist strong
homotopy $F,$ such that $F_0(\mathcal{B})=\iota,$ $F_1(\mathcal{B})=\iota'.$
Homotopy relation is equivalence on $\mathcal{E}$ and $\hm{\mathcal{E}}
 \equiv\mathcal{E}/\hmt{}{}\equiv\pi_H(\mathcal{E}).$

Lets define strong homotopic equivalence on  $\tm{\mathcal{E}}{\mathcal{E}}.$
We' ll say, that $\hmt{\de{1}{2}}{\depr{1}{2}},$ if simultaneously
$\hmt{\iota_1}{\iota'_1}$ и $\hmt{\iota_2}{\iota'_2}.$ Evidently, the set of
classes of the strong homotopic equivalence
$\hm{\tm{\mathcal{E}}{\mathcal{E}}}=\pi_H(\tm{\mathcal{E}}{\mathcal{E}})$
coincides with
$\tm{\hm{\mathcal{E}}}{\hm{\mathcal{E}}}=\tm{\pi_H(\mathcal{E})}{\pi_H(\mathcal{E})}.$

Now we are able to define some special kinds of deformations in $\df.$ Lets
consider the set $\pi^{-1}_H(\Delta(\tm{\hm{\mathcal{E}}}{\hm{\mathcal{E}}})),$
i.e. the set of pair of embeddings, that are homotopic to each other. The set,
after factorization by $\pi_D$ induces the subset $\df_0\subseteq\df,$ which
we'll call {\it proper deformation.} With\-in the classical (nonquantum) theory
of deformational structures we'll be concerning only with this type of
deformations. Evidently,
 $\df_0$
---  subpseudogroup of  $\df.$

Lets consider the set of pair from
$\pi_D\circ\pi^{-1}_H(\Delta(\tm{\hm{\mathcal{E}}}{\hm{\mathcal{E}}}))$ of the
kind $d(\iota)=[(\iota,\iota\circ l_0)]_D,$ where $\iota\in\mathcal{E},\
l_0\in[{\rm id}_{\mathcal{B}}]_H.$ Evidently, that the set $\{d(\iota)\}$ is no
more, less then smooth proper diffeomorphisms $\Diff{\mathcal{S}}_0$ of the
deformant. We'll call it {\it proper sliding} of the deformant
$\mathcal{S}=\iota(\mathcal{B})$ in $\mathcal{M}$ and denote $\mathrm{S}{\rm
l}_{\iota}.$ Evidently, $\mathrm{S}{\rm l}_\iota$ is subpseudogroup of $\df_0,$
isomorphic to the group $\pi^{-1}_H[{\rm id}_\mathcal{B}].$

\section{Vector fields, rigid motions, generalized Killing equations
and $d-$coverings of $d-$manifolds.}

Lets consider some proper deformation $\zeta_0=[(\iota,\iota')]_D\in\df_0:\
\mathcal{S}=\iota(\mathcal{B})\to\mathcal{S'}=\iota'(\mathcal{B}).$ Let $F_t$
--- some homotopy, connecting  $\iota$ and $\iota'.$ Consider the set
$\mathcal{M}\supset\mathcal{P}_{\mathcal{S}\mathcal{S'}}=\cup_{t\in
I}F_t(\mathcal{B})\equiv\cup_{t\in I}\mathcal{S}_t.$ It can be treated as
smooth mapping of the smooth manifold $I\times\mathcal{B}\to \mathcal{M},$
which, generally speaking, is not submanifold and even is not immersion in
$\mathcal{M}.$ Its boundary $\partial\mathcal{P}_{\mathcal{S}\mathcal{S'}}$ is
$\mathcal{S}\cup\mathcal{S'}\cup_{t\in I }F_t(\partial\mathcal{B}).$ Let
$\widetilde{d/dt}$ ---  uniquely determined horizontal vector field on
$\tm{\mathcal{B}}{I},$ i.e. such that $d\pi_1(\widetilde{d/dt})=0,$
$d\pi_2(\widetilde{d/dt})=d/dt,$ where $\pi_1,\pi_2$ --- projections of
$\tm{\mathcal{B}}{I}$ to $\mathcal{B}$ and $I$ correspondingly. The set
$\mathcal{P}_{\mathcal{S}\mathcal{S'}}$ is composed of an integral lines
$\{F_t(b)\}_{b\in\mathcal{B}}$ of the vector field $v=dF(\widetilde{d/dt}),$
defined on $\mathcal{P}_{\mathcal{S}\mathcal{S'}}.$ The family of embeddings
$\{F_t(\mathcal{B})\}_{t\in I}$ induces the family of deformation forms
$\{\Delta^t_\mathcal{B}\}_{t\in I}$ by the following rule:
$$\Delta^t_\mathcal{B}=(dF_t)^\ast\Theta-(dF_0)^\ast\Theta.$$ We'll say, that
the homotopy $F_t$ is {\it rigid motion} of the $d-$body $\mathcal{B}$ in
$\mathcal{M}$ from the configuration $S=F_0(\mathcal{B})$ to the configuration
$\mathcal{S}'=F_1(\mathcal{B}),$ if $\Delta^t_{\mathcal{B}}=0$ for any $t\in I$
or, in other words the image $(dF_t)^\ast\left.\Theta\right|_{\mathcal{S}_t}$
is constant on $I$.

\begin{pred}
Homotopy $F_t$ from  $\mathcal{S}$ to $\mathcal{S}'$ is rigid motion if and
only if
\begin{equation}\label{killing}
\mathrm{L}_v\left.\Theta\right|_{\mathcal{P}}=0,
\end{equation}
where $\mathrm{L}_v$ --- Lie derivative along the vector field
$v=d_tF_t(\mathcal{B},d/dt)$ on $\mathcal{P}_{\mathcal{S}\mathcal{S}'}.$
\end{pred}
The equations of the type (\ref{killing}) we'll call {\it generalized Killing
equations.}

Let $\mathcal{S}=\iota(\mathcal{B})$ will be some fixed deformant  and let
$\mt{\mathcal{S}}$ --- set of its all possible rigid motions in $\mathcal{M}.$
The set is, generally speaking, proper subset of path connected component of
the embedding  $\iota$  in $\mathcal{E}$ (the component is exactly the class
$[\iota]_H$ of homotopic to $\iota$ embeddings), which is defined by the
specification of a $d-$metrics. Easily to see, that rigid motions define
equivalency $\stackrel{M}{\sim}$ on $\mathcal{E}$: we'll call the two
embeddings $\iota$ и $\iota'$ --- equivalent: $\iota\stackrel{M}{\sim}\iota',$
if there exist homotopy $F\in\mt{\iota(\mathcal{B})},$ connecting $\iota$ and
$\iota'.$ Obviously, the equivalency $\stackrel{M}{\sim}$ is more weak then
$\stackrel{H}{\sim},$ then the class $[\iota]_H=\cup_\alpha[\iota_\alpha]_M,$
where $\{\iota_\alpha\}$ --- some set of all pairwise
$\stackrel{M}{\sim}$-nonequivalent elements from $[\iota]_H,$ and
$[\iota_\alpha]_M\cap[\iota_\beta]_M=\varnothing$ for all $\alpha\neq\beta.$
We'll call $[\iota_\alpha]_M$ --- {\it $\alpha$-component} $[\iota]_H,$ and its
image $$\mathcal{P}_{\mathcal{S}_\alpha}\equiv\bigcup\limits_{F\in
\mt{\mathcal{S}_\alpha}}\mathcal{P}_{\mathcal{S}_\alpha F(\mathcal{B})}$$ {\it
rigidity $\alpha-$component of the manifold $\mathcal{M}$ relatively to
embedding $\iota.$} The family $\{\mathcal{P}_{\mathcal{S}_\alpha}\}$ gives
some covering of $\mathcal{M}$:
$$\mathcal{M}=\bigcup\limits_{\alpha}\mathcal{P}_{\mathcal{S}_\alpha},$$ which
we'll call {\it deformational $(\mathcal{B},\Theta,h)-$co\-ve\-ring of the
manifold $\mathcal{M},$} where $\hm{\mathcal{E}}\ni h=\pi_H(\iota)$, or, more
shortly, {\it $d-$covering.}

Within the theory of dynamical deformable structures,  where physical action
should be considered as functional of deformation form:
$\mathcal{A}=\mathcal{A}[\Delta],$ it is naturally to use as configuration
space of deformant not $[\iota(\mathcal{B})]_H,$ but its factor:
$$\{[\iota(\mathcal{B})]_H/\stackrel{M}{\sim}\}\equiv\pi_{M}([\iota(\mathcal{B})]_H)
\simeq\{\mathcal{P}_{\mathcal{S}_\alpha}\},$$ which reflects the deformational
indistinguishability  of those configurations, that are connected by a some
rigid motion. We'll call the manifold $\mathcal{M}$ {\it deformationally
discrete} relatively to its rigid $(\mathcal{B},\Theta,h)-$co\-ve\-ring, if
$\pi_M$
--- identical mapping, and {\it deformationally trivial,} if $\pi_M$
--- constant mapping. Then, the manifold $\mathcal{M}$ will be called
{\it deformationally homogeneous ($d-$homogeneous),} if
\begin{equation}\label{hom}
\mathcal{P}_{\mathcal{S}_\alpha}=\mathcal{M}
\end{equation}
for some  $\alpha$ and  {\it completely deformationally homogeneous,} if
(\ref{hom}) is valid for all $\alpha.$

Deformationally trivial manifolds, have no significance from the view point of
deformational structure theory, by the following
\begin{pred}
Any deformationally trivial manifold has zero $d-$metrics.
\end{pred}
Riemannian manifold with general metrics $g$ is an example of deformationally
discrete manifold. The euclidian space $\mathbb{E}^n$ is completely
deformationally homogeneous relatively $(\mathcal{B},\eta,h)-$decomposition,
whe\-re $\eta$ --- euclidian metric, $\mathcal{B}$ --- arbitrary $d-$body, $h$
--- arbitrary element $\hm{\mathcal{E}}.$
As an example of deformationally homogeneous but not completely deformationally
homogeneous manifolds lets consider the following situation. Let
$\mathcal{M}=\overline{D^2}_{2r}(0)\setminus D^2_r(0)$ --- closed ring on 2D
euclidian plane (as usually, $D^n_r(a)$ --- $n-$dimensional disk with radius
$r$ and center $a,$ bar above letter - closure), $\Theta=\eta$
--- 2D euclidian metrics, $\mathcal{B}=S^1,$ $\iota(S^1)=S'^1_{R}\subset\mathcal{M}$ ---
circle with radius $R$ and $\pi_H(\iota)=1$ (in the considered case
$\pi_H(\mathcal{E})\equiv\pi_1(\mathcal{M})$
--- fundamental group of $\mathcal{M},$ isomorphic $\mathbb{Z}.$) Then, in case
$R<3r/2,$  $\mathcal{P}_{S^1_R}=\overline{D^2}_{2R-r}(0)\setminus
D^2_r(0)\neq\mathcal{M}$ and only in case  $R=3r/2$ we have
$\mathcal{P}_{S^1_{3r/2}}=\mathcal{M}.$

In conclusion we formulate two propositions and give an example, all
illustrating more general character of the deformational structure theory in
comparison with isometry theory of Riemannian spaces.

Let ${\rm Imm}(v)\equiv\{p\in\mathcal{M}\,|\ \phi_t(p)=p\}$ --- is the set of
all immobile points of the full one-parametric group $\phi_t$, generated by
some smooth vector field $v.$

\begin{pred}\ If manifold $\mathcal{M}$ admits isometry of $d-$metrics,
i.e. if there exists vector field $v\in T\mathcal{M},$ such that
$\mathrm{L}_v\Theta=0,$ then $\forall\ \mathcal{S}$ such that
$\mathcal{S}\not\subseteq{\rm Imm}(v),$ there exists nonidentical rigid motion
$\{\left.\phi_t\right|_{\mathcal{S}}\}\in$\\ $\mt{\mathcal{S}}$ and, by the
fact, $\mathcal{M}$ is not deformationally discrete.
\end{pred}

\begin{pred}
If manifold $\mathcal{M}$ admits $r-$parametric isometry group $\mathcal{G}$,
generated by  vector fields $\left\{v_1,\dots\right.$ $\left.,v_r\right\},$
such that $\mathrm{L}_{v_i}\Theta=0,\ i=1,\dots,r,$ that act on $\mathcal{M}$

1) transitively, then $\mathcal{M}$ --- completely deformationally homogeneous
(relatively any decomposition) ;

2) intransitively, and if also $\mathcal{S}\cap{\rm Orb}\, \mathcal{G}$
--- connected for some orbit ${\rm Orb}\,\mathcal{G}\subset\mathcal{M}$, then
${\rm Orb}\,\mathcal{G}$ --- completely deformationally homogeneous relatively
its $(\iota^{-1}(\mathcal{S}\cap{\rm
Orb}\,\mathcal{G}),\left.\Theta\right|_{{\rm Orb}\,\mathcal{G}},
h=\pi_{H}(\iota))$-decomposition. Here, as usually,
$\mathcal{S}=\iota(\mathcal{B}).$
\end{pred}
Particularly, if $\mathcal{S}=\iota(\mathcal{B})={\rm Orb}\,\mathcal{G},$ then
$\mt{\mathcal{S}}\cap{\rm Sl}_\iota\neq\varnothing$ defines the group of {\it
rigid proper sliding.}

So, if $\Theta$ --- Riemannian (or any other $d-$) metrics on  $\mathcal{M}$
and $\mathcal{M}$ admits isometry,  then nontrivial rigid motions of
$d-$objects will be always exist. The following example shows, that inverse is
not valid.

Let $\mathcal{M}=\mathbb{R}^2$ with cartesian coordinate system $\{x_1,x_2\},$
$\mathcal{B}=I=[0,1]\in\mathbb{R},$
$\Lambda^1(\mathbb{R}^2)\ni\Theta=(x^1x^2+\coth x^2)dx^1.$ Let
$\iota(\mathcal{B})\equiv\mathcal{S}=\{0\le x^1\le 1,\,x^2=0\}.$ By the fact,
that $\left.\Theta\right|_{x^2=0}=dx^1={\rm const},$ it is easily to see that
the set of homotopies $$\{F_t: \mathcal{S}\to\mathcal{S}_t=(x_1+t,0),\, 0\le
x_1\le1, -\infty<t<\infty\},$$ (they are simple rigid translations of units
interval along axe $x^1$) lies in $\mt{\mathcal{S}}.$ Moreover,
$\mathcal{P}_{\mathcal{S}}=\mathbb{R}^1=\{(x^1,0)\}.$ The related vector field
$v(t,x^1),$ along which
 $\mathrm{L}\left.\Theta\right|_{\mathcal{P}}=0$ is simply $\partial/\partial x^1.$
It is easily to show, that $v$ does'nt admit smooth continuation $\tilde v$
from the $\mathcal{P}_{\mathcal{S}}\subset\mathbb{R}^2$ to $\mathbb{R}^2.$
Really, Killing equations  $\mathrm{L}_{\tilde v}\Theta=0$ for this case with
account $\left.\tilde v\right|_{\mathcal{P}_{\mathcal{S}}}=\partial/\partial
 x^1$ reads:
 $$\tilde v^2=-\frac{x^2}{x^1+\sinh x^2}.$$ The component has singularity
 on line $x^1=-\sinh x^2,$ which cross any neighborhood of
 $\mathcal{P}_{\mathcal{S}}$ in $\mathbb{R}^2.$

\vspace{1cm} {\bf Acknowledgements}\\
I am very thankful  to local organizers
of the 5-th Asian-Pacific conference
 for comfortable work conditions and full financial supporting, and also to
the professor of algebra chair of YSPU A.S.Tikhomirov for valuable
consultations during the preparing of the paper.


\small
\begin{thebibliography}{99}

\bibitem{pav}
M.~Pav\u{s}i\u{c},~V.~Tapia, ``Resource letter on geometrical results for
embeddings and branes''. gr-qc/0010045.
\bibitem{green}
M. Green, J. Shwarz, E. Witten, ``Superstring theory'', Cambridge University
Press, Cambridge, 1989.
\bibitem{plectures}
C.~P.~Bachas, ``Lectures on D-branes''. hep-th/9806199
\bibitem{noncomm}
I.~Ya.~Aref'eva~and~others, ``Noncommutative Field Theories and (Super)String
Field Theories''. hep-th/0111208
\bibitem{kobayasi}
Sh.~Kobayashi,~K.~Nomizu, ``Foundations of differential geometry'', v.1,2,
NFMI, 1999 (In Russian).
\bibitem{vinogr}
A.M.Vinogradov,~I.S.Krasilshchik and others,``Symmetries and conservation laws
of equations of mathematical physics'', M., Factorial, 1997 (In Russian).
\bibitem{land}
L. D. Landau, E. M. Lifshits, ``Elasticity theory'',  Moscow, Nauka, 1987, (In
Russian).
\bibitem{saharov}
A. D. Sacharov, DAN USSR {\bf 177}, 70, (1967) (in Russian).
\bibitem{tar}
A.~Tartaglia, {\it Grav. \& Cosmol.} {\bf 1}, 335 (1995).
\bibitem{hehl}
V.~F.~Hehl,~I.~Newman, {\it In} pr.``Perspectivy edinoy teorii'',  MSU,
p.137-166, 1991 (In Russian); {\it also}  F. Gronwald and F. Hehl,
gr-qc/9701054.
\bibitem{dmitriev}
V. A. Dmitriev, ``Stochastic mechanics'', Moscow, Vysshaya shcola, 1990, (In
Russian).
\bibitem{kok1}
S.~S.~Kokarev, {\it Grav. \& Cosmol.} {\bf 2}, 96 (1998).
\bibitem{kok2}
S.~S.~Kokarev, {\it Nuovo Cimento B} {\bf 113},  1339 (1998).
\bibitem{kok3}
S.~S.~Kokarev, {\it Nuovo Cimento B} {\bf 114},  903 (1999).
\bibitem{kok4}
S.~S.~Kokarev,  {\it Nuovo Cimento B} {\bf 116}, 915 (2001).
\bibitem{warner}
F.Warner, ``Differentiable manifolds''.
\bibitem{trus}
C. Truesdell, ``A first course in  rational continuum mechanics'', The Johns
Hopkins University, Baltimore, Maryland, 1972.
\end{thebibliography}
\end{document}